\documentclass[reprint,superscriptaddress,amsmath,amssymb,aps,pre]{revtex4-2}
\usepackage{graphicx}
\usepackage{dcolumn}
\usepackage{bm}
\usepackage{color}
\usepackage{units}
\usepackage{wasysym}
\usepackage{MnSymbol}
\usepackage{comment}
\usepackage{times}
\usepackage[unicode=true,pdfusetitle,bookmarks=false,colorlinks=true,citecolor=blue,urlcolor=blue,linkcolor=blue]{hyperref}
\begin{document}

\title{Tuning Nonequilibrium Phase Transitions with Inertia}
\author{Ahmad K. Omar}
\email{aomar@berkeley.edu}
\affiliation{Department of Materials Science and Engineering, University of California, Berkeley, California 94720, USA}
\affiliation{Materials Sciences Division, Lawrence Berkeley National Laboratory, Berkeley, California 94720, USA}
\author{Katherine Klymko}
\email{kklymko@lbl.gov}
\affiliation{NERSC, Lawrence Berkeley National Laboratory, Berkeley, California 94720, USA}
\affiliation{Computational Research Division, Lawrence Berkeley National Laboratory, Berkeley, California 94720, USA}
\author{Trevor GrandPre}
\affiliation{Department of Physics, University of California, Berkeley, California 94720, USA}
\author{Phillip L. Geissler}
\affiliation{Department of Chemistry, University of California, Berkeley, California 94720, USA}
\affiliation{Chemical Sciences Division, Lawrence Berkeley National Laboratory, Berkeley, California 94720, USA}
\author{John F. Brady}
\affiliation{Division of Chemistry and Chemical Engineering, California Institute of Technology, Pasadena, California 91125, USA}

\begin{abstract}
In striking contrast to equilibrium systems, inertia can profoundly alter the structure of active systems.
Here, we demonstrate that driven systems can exhibit effective equilibrium-like states with increasing particle inertia, despite rigorously violating the fluctuation-dissipation theorem.
Increasing inertia progressively eliminates motility-induced phase separation and restores equilibrium crystallization for active Brownian spheres.
This effect appears to be general for a wide class of active systems, including those driven by deterministic time-dependent external fields, whose nonequilibrium patterns ultimately disappear with increasing inertia. 
The path to this effective equilibrium limit can be complex, with finite inertia sometimes acting to accentuate nonequilibrium transitions.
The restoration of near equilibrium statistics can be understood through the conversion of active momentum sources to passive-like stresses.
Unlike truly equilibrium systems, the effective temperature is now density dependent, the only remnant of the nonequilibrium dynamics.
This density-dependent temperature can in principle introduce departures from equilibrium expectations, particularly in response to strong gradients.
Our results provide additional insight into the effective temperature ansatz while revealing a mechanism to tune nonequilibrium phase transitions.

\end{abstract}

\maketitle

\textit{Introduction.--}
Particle dynamics that break time-reversal symmetry result in a non-Boltzmann distribution of microstates, leading to phase transitions and pattern formation that defy equilibrium intuition.
Self-propelled particles can phase separate in the absence of cohesive interactions, a phenomenon commonly referred to as motility-induced phase separation (MIPS)~\cite{Fily2012, Redner2013, Buttinoni2013, Cates2015}.
Externally manipulating the trajectories of particles (e.g.,~with magnetic or electric fields) can result in microphases and pattern formation~\cite{Dzubiella2002, Vissers2011, Klymko2016, Han2017, delJunco2018, DelJunco2019} despite purely repulsive, simple interaction potentials.
While the observation of these and numerous other nonequilibrium phase transitions in natural and synthetic systems is now routine, the theoretical description of these transitions is clouded by the absence of \textit{a priori} knowledge of the distribution of microstates.

Understanding the many-body phase behavior of driven systems remains a principal challenge in nonequilibrium statistical mechanics.
However, it has become clear that some nonequilibrium systems may admit an \textit{effective} Boltzmann distribution of states, a feature that radically improves our ability to understand these systems~\cite{Cugliandolo2011}.
Athermal granular materials have been described by the Edwards ensemble, which, in simple terms, is a statistical mechanical framework based on the assertion that states with equal volume (rather than energy) have equal probability~\cite{Edwards1989, Mehta1989}.
The fluctuations of boundary-driven arrested materials have also been posited to be controlled by an effective temperature (albeit, not uniquely defined)~\cite{Ono2002, OHern2004}.

In the case of active or locally driven matter -- where the breaking of time-reversal symmetry is due to a particle-level driving force -- effective equilibrium states have been invoked to explain emergent structural features and phase transitions with varying degrees of success~\cite{Farage2015, Takatori2015, Rein2016, Wittmann2016, Wittmann2017a, Wittmann2017b, Han2017, Wittmann2019, Turci2021}. 
In these scenarios, energy scales born from the active dynamics (which can be defined using, e.g.,~linear response theory or invoking Stokes-Einstein relations) play the role of an effective temperature that is then used to map these driven systems onto equilibrium analogues~\cite{Loi2008, Palacci2010, Wang2011, Loi2011, Morozov2010, Szamel2014, Solon2015b, Takatori2016, GrandPre2018, Burkholder2019}.
However, assessing the success of effective temperature ideas to describe nonequilibrium steady states remains largely empirical.
Recent work by O'Bryne and Tailleur demonstrated that the dynamics of tactic active matter can be rigorously mapped to effective passive systems at a hydrodynamic level~\cite{OByrne2020, OBryne2021}.
On a microscopic level, active forces under certain limits can act as traditional thermal forces by satisfying an effective fluctuation-dissipation theorem (FDT)~\cite{Fodor2016, Mandal2017}.
However, the general question remains: When can nonequilibrium systems be rigorously mapped to effective equilibrium states? 

\begin{figure*}
	\centering
	\includegraphics[width=.95\textwidth]{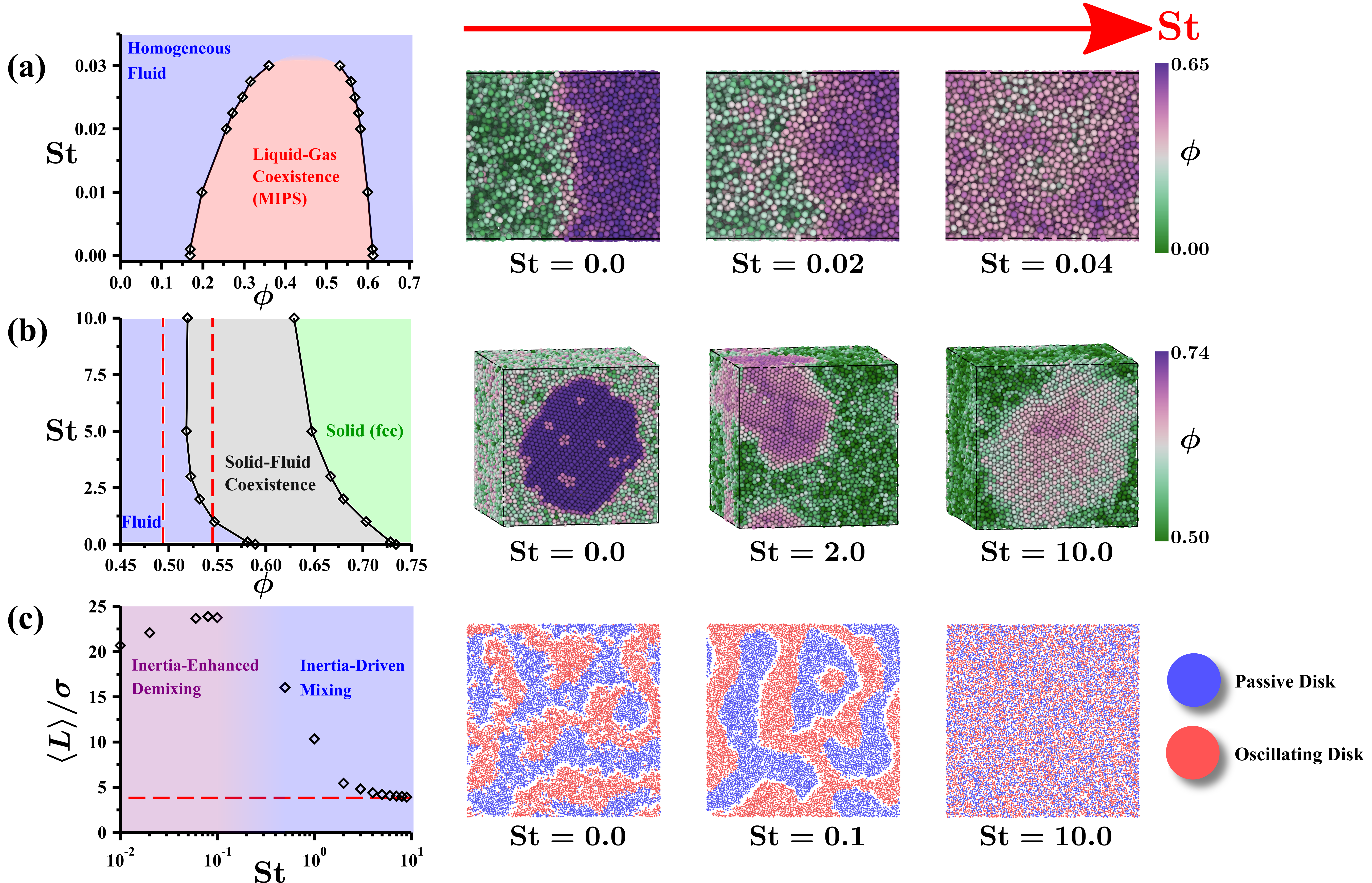}
	\caption{\protect\small{{Inertia dependence of (a) MIPS ($\ell_0/\sigma =50.0$) and (b) crystallization ($\ell_0/\sigma =5.0$) of active Brownian spheres (see Ref.~\cite{Omar2021} for ${\rm St} = 0$ phase diagram) and (c) pattern formation in a $50:50$ mixture (with number density $\rho \sigma^2 = 0.45$) of oscillating ($U_0 \sigma/ D^B = 70, \ell_0/\sigma \equiv U_0/\omega = 1.67$) and passive soft disks (see Ref.~\cite{delJunco2018} for ${\rm St} = 0$ phase diagram). Dashed lines denote equilibrium expectations.}}}
	\label{figure1}
\end{figure*}

In this Article, we explore the question raised above in the context of active systems.
Despite the presence of active forces that strictly violate the FDT on a microscopic level, we demonstrate a new limit in which nominally nonequilibrium systems can be pushed to effective equilibrium states simply by increasing inertia. 
This limit occurs when the translational momentum relaxation time $\tau_M$ is much larger than the intrinsic timescale associated with translational active forces $\tau_A$, as measured through the Stokes number ${\rm St} \equiv \tau_M/\tau_A$.
When ${\rm St} \rightarrow \infty$, active forces behave as thermal forces with the kinetic temperature playing the role of a now \textit{density-dependent} effective temperature.
We show that a variety of nonequilibrium transitions observed in the overdamped limit are attenuated with increasing inertia while equilibrium-like transitions are restored. 
Our results offer a new interpretation for the reported~\cite{Mandal2019, Lei2019, Dai2020, Liao2021, Negro2022} dependency of nonequilibrium phase transitions on translational inertia and further insight into the applicability of the effective temperature perspective.

\textit{Many-Body Phase Behavior.--}
While the impact of inertia on active dynamics has been recently investigated~\cite{Attanasi2014, Manacorda2017, Takatori2017, Scholz2018, Mandal2019, Lowen2020, Sandoval2020, Dai2020, Liao2021, Caprini2021, Su2021, Sprenger2021, Chatterjee2021, teVrugt2021, Negro2022, Martins2022, teVrugt2022}, a generalized understanding of its impact on active phase behavior has remained elusive. 
Before proceeding to demonstrate the impact of inertia on the many-body phase behavior of driven systems, we briefly discuss the model systems and their known equilibrium limit.
We consider the microscopic distribution of the positions $\mathbf{x}^N$ of $N$ particles each with mass $m$ and each experiencing conservative interparticle/external forces $\mathbf{F^C}=\mathbf{F^{\rm int}} + \mathbf{F^{\rm ext}} = -\bm{\nabla}V(\mathbf{x}^N)$ (where $V$ is the potential) and two nonconservative forces: a drag force $\mathbf{F^{\rm drag}}$ that dissipates the particle's energy and a fluctuating source force $\mathbf{F^{\rm source}}$ that injects energy into the system. 
The FDT establishes the relationship between the fluctuating and dissipative forces that is required to rigorously recover a Boltzmann distribution of microstates $\mathcal{P}(\mathbf{x}^N) \propto \exp[-V(\mathbf{x}^N)/k_BT]$ where $k_BT$ is the energy scale associated with the source.  
Taking the drag coefficient of each particle to be independent, memoryless, and constant (i.e.~$\mathbf{F^{\rm drag}} = -\zeta \mathbf{u}$ where $\zeta$ is the drag coefficient and $\mathbf{u}$ is the particle velocity), the FDT constrains the source force to have a mean of $\mathbf{0}$ and a variance of $\langle \mathbf{F^{\rm source}}(t)\mathbf{F^{\rm source}}(t') \rangle = 2k_BT\zeta \delta(t-t')\mathbf{I}$. 

Nonequilibrium forces can satisfy the FDT in certain limits, making the notion of an effective temperature exact under these conditions. 
For instance, for active Brownian particles (ABPs), the source of fluctuations is $\mathbf{F^{\rm source}} = \mathbf{F^{\rm act}} = \zeta U_0 \mathbf{q}$ where $U_0$ is the intrinsic active speed and $\mathbf{q}$ is the particle orientation. 
Taking the dynamics of $\mathbf{q}$ to be diffusive and overdamped with rotary diffusion constant $D_R$, the particles move persistently for a characteristic timescale $\tau_R = D_R^{-1}$. 
Rotational inertia can profoundly alter the dynamics of ABPs (see the recent work of Sandoval~\cite{Sandoval2020}) we neglect such effects here to simplify our analysis and isolate the role of translational inertia on systems with translational active forces.
While generally not satisfying the FDT, $\mathbf{F^{\rm act}}$ does so in the limit that $\tau_R \rightarrow 0$ (i.e.,~$\langle \mathbf{F^{\rm act}}(t)\mathbf{F^{\rm act}}(t') \rangle = 2k_BT^{\rm eff}\zeta \delta(t-t')\mathbf{I}$) with an effective temperature of $k_BT^{\rm eff} = \zeta D^{\rm act}$ where $D^{\rm act} = \frac{U_0^2 \tau_R}{d(d-1)}$ is the intrinsic active diffusion constant in $d$ spatial dimensions~\cite{Mandal2017, Omar2021}. 
This is often the limit explored in experiments that have observed effective Boltzmann distributions in active systems~\cite{Palacci2010, Takatori2016}.

Short of this limit, with the microscopic dynamics in strict violation of the FDT, we now demonstrate the implications of inertia on the phase behavior of active systems.
We conduct simulations~\cite{Plimpton1995, Anderson2020} (see Supplemental Material (SM) for details~\cite{Note1}) of effective hard-sphere ABPs in three dimensions (3D), the phase behavior of which has been recently established~\cite{Turci2021, Omar2021}.
In the overdamped hard-sphere limit, the phase diagram is characterized by two geometric parameters: the ratio of the run length $\ell_0\equiv U_0\tau_R$ to particle diameter $\sigma$ and the volume fraction $\phi$.  
Here, the system undergoes MIPS for a broad range of $\ell_0/\sigma$ and $\phi$~\cite{Omar2021}. 
The addition of translational inertia adds a third dimensionless axis to the phase diagram, quantified by ${\rm St} \equiv \tau_M/\tau_A \equiv (m/\zeta)/\tau_R$.
Selecting a state well within the regime of MIPS such that the liquid and gas densities are appreciably distinct, we observe [see Fig.~\ref{figure1}(a)] a rapid and monotonic elimination of coexistence with increasing ${\rm St}$ and the absence of MIPS entirely when ${\rm St} > 0.03$. 

The sensitivity of MIPS to inertia in 2D was first reported by Mandal~\textit{et al.}~\cite{Mandal2019} who demonstrated that, for all values of $\ell_0/\sigma$ and for high enough ${\rm St}$, a homogeneous state was observed in lieu of MIPS. 
As we later argue, the absence of MIPS with increasing inertia is rooted in the system reaching an effective equilibrium state.
If this is indeed the case, we should additionally observe the restoration of equilibrium transitions.
In the case of our hard-sphere system, Boltzmann statistics would result in the equilibrium freezing transition with coexisting fluid and solid densities~\cite{Hoover1968, Pusey2009} of $\phi_{\rm fluid}=0.494$ and $\phi_{\rm solid}=0.545$.
Importantly, equilibrium freezing is athermal in origin and thus does not depend on the precise value of $k_BT^{\rm eff}$.

In the overdamped limit, activity dramatically shifts the freezing transition towards higher densities~\cite{Omar2021} with the solid phase exhibiting a nearly close-packed density ($\phi_{\rm solid} \rightarrow 0.74$) for activities as small as $\ell_0 / \sigma =5.0$ [see $\rm{St} = 0$ in Fig.~\ref{figure1}(b)].
In the limit of small run lengths $\ell_0/\sigma\rightarrow 0$ where FDT is satisfied, it was recently shown that, despite the use of a continuous potential, athermal active particles display a freezing transition that closely resembles that of equilibrium hard spheres~\cite{Omar2021}. 
Departing from the overdamped limit, we observe a shift in the crystal coexistence region to lower densities [see Fig.~\ref{figure1}(b)], consistent with our expectations of a return to an effective equilibrium. 
We note that the crystallization transition appears to be much less sensitive to inertia than MIPS, with the coexistence window continuing to shift to lower densities for $\rm{St} > 1$. 
For ${\rm St} > 5$, the densities (particularly that of the fluid) appear to saturate at values that, while significantly closer, are decidedly distinct from the expected values of equilibrium freezing.
The source of this discrepancy are possibly rooted in the nonequilibrium origins of the effective temperature, which we later show manifest in its \textit{density dependence}.

Thus far, increasing inertia has monotonically pushed the phase behavior of active systems to more closely resemble effective equilibrium systems. 
However, it should be emphasized that reaching the effective equilibrium appears to occur in the asymptotic limit of high inertia.
Intermediate values of inertia can also \textit{heighten} the nonequilibrium features observed in overdamped phase transitions.
2D systems of repulsive active-passive mixtures are known to form microphases when the source of activity is an oscillatory deterministic force ($\mathbf{F^{\rm act}} = \zeta U_0 \mathbf{q}$ with $\mathbf{q}(t) = \sin(\omega t)\mathbf{\hat{e}_x} + \cos(\omega t)\mathbf{\hat{e}_y}$)~\cite{Han2017, delJunco2018}. 
Here, all particles experience Brownian forces $\mathbf{F^B}$ (resulting in a translational diffusivity $D^B\equiv k_BT/\zeta$) that do satisfy the FDT (i.e.,~$\mathbf{F^{\rm source}} = \mathbf{F^B}$ and $\mathbf{F^{\rm source}} = \mathbf{F^{\rm act}} + \mathbf{F^B}$ for the passive and active particles, respectively).  
By introducing translational inertia with ${\rm St} \equiv \tau_M/\tau_A \equiv (m/\zeta)\omega$, we observe [see Fig.~\ref{figure1}(c)] an enhancement in the average size $\langle L \rangle$ of the microdomains (and interfacial smoothness) with increasing inertia before eventually reaching the effective equilibrium limit.
That there exists an optimum ${\rm St}$ for accentuating nonequilibrium transitions makes clear the path towards effective equilibrium can be complex. 

\textit{Theory and Discussion.--}
We now explore the origins of the apparent return to equilibrium phase behavior of active systems with increasing inertia.
Before proceeding to interacting systems it is instructive to consider the distribution of ideal ABPs ($\mathbf{F^{\rm int}} = \mathbf{0}$), which in addition to the active force, also experience translational Brownian forces $\mathbf{F^B}$. 
These dynamics result in a Fokker-Planck equation for the probability density $\mathcal{P}(\mathbf{x}, \mathbf{u}, \mathbf{q},t)$~\cite{Risken1989}:
\begin{equation}
\label{eq:idealfp}
\frac{\partial\mathcal{P}}{\partial t} = -\bm{\nabla}_{\mathbf{x}}\cdot\mathbf{j^x} -\bm{\nabla}_{\mathbf{u}}\cdot\mathbf{j^u}  -\bm{\nabla}_{\mathbf{q}}\cdot\mathbf{j^q},
\end{equation}
where $\mathbf{j}^{\bm{\psi}}$ and $\bm{\nabla_\psi}$ are the flux of probability and gradient operator in $\bm{\psi}$-space, respectively (see SM~\cite{Note1}). 

\begin{figure*}
	\centering
	\includegraphics[width=.65\textwidth]{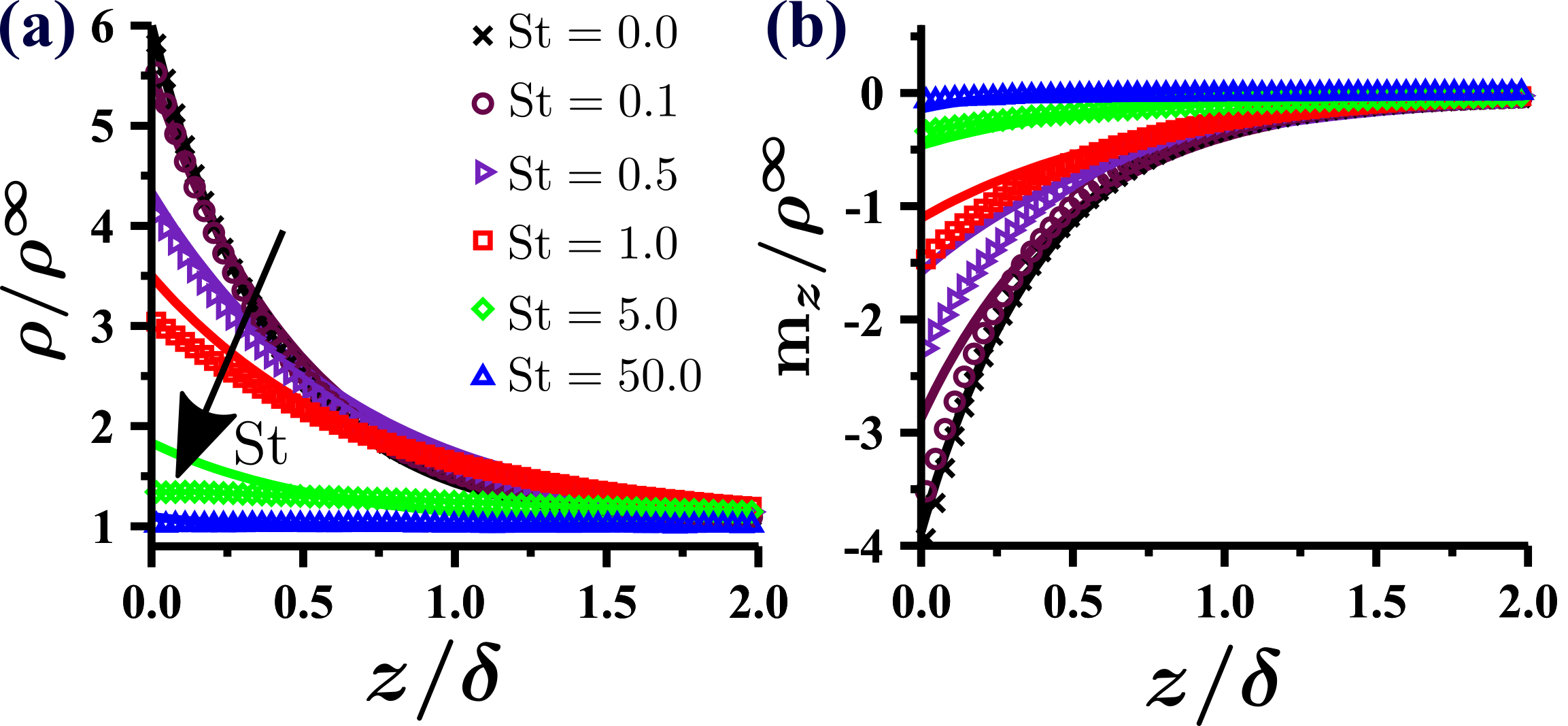}
	\caption{\protect\small{{Simulation (symbols) and theoretical (lines) results of (a) $\rho$ and (b) ${\rm m}_z$ [Eq.~\eqref{eq:polarorderprofile}] for ideal ABPs with $D^{\rm act}/D^B=5.0$ near a no-flux boundary. }}}
	\label{figure2}
\end{figure*}

By taking statistical moments~\cite{Note1} of Eq.~\eqref{eq:idealfp} in both $\mathbf{u}$ and $\mathbf{q}$, one finds that the steady-state number density field $\rho(\mathbf{x}) = \int \mathcal{P}\,d\mathbf{u}\,d\mathbf{q}$ of ideal ABPs satisfies mass conservation $\bm{\nabla}_{\mathbf{x}} \cdot \mathbf{j^\rho} = 0$ where the density flux $\mathbf{j^\rho}$ can be determined from linear momentum conservation:
\begin{equation}
\label{eq:momentumbalance}
\bm{\nabla}_{\mathbf{x}}\cdot \underbrace{\bm{\sigma}}_{\substack{\text{particle} \\ {\text{stress}}}}  + \underbrace{\zeta U_0 \mathbf{m}}_{\substack{\text{active body} \\ {\text{force}}}} + \underbrace{\rho\mathbf{F^{\rm ext}}}_{\substack{\text{external} \\ {\text{body force}}}} -\underbrace{\zeta\mathbf{j^\rho}}_{\substack{\text{drag body} \\ {\text{force}}}} = \mathbf{0},
\end{equation}
where $\mathbf{m}(\mathbf{x}) = \int \mathcal{P}\,\mathbf{q}\,d\mathbf{u}\, d\mathbf{q}$ is the polarization density of the particles.
For ideal ABPs, the stress is simply the kinetic stress $\bm{\sigma} = \bm{\sigma^K} = -\rho m \langle \mathbf{uu} \rangle$.
The nonconservative and external forces act as body forces -- sources or sinks of momentum that can generate stress gradients in a system.  
The active body force~\cite{Yan2015a, Rodenburg2017, Epstein2019, Omar2020, Omar2022} injects momentum into the system through the polarization field which, at steady state, satisfies $\mathbf{m} = -\frac{\tau_R}{d-1}\nabla_\mathbf{x} \cdot \mathbf{j^m}$.
An active (or ``swim'') stress~\cite{Mallory2014, Fily2014, Takatori2014, Solon2015, Solon2015a} is often defined $\bm{\sigma^{\rm act}}=-\frac{\zeta U_0 \tau_R}{d-1} \mathbf{j^m} = -\frac{\zeta U_0 \tau_R}{d-1} \rho \langle \mathbf{uq} \rangle\sim\tau_R\rho\langle \mathbf{u}\mathbf{F^{\rm act}}\rangle$, allowing Eq.~\eqref{eq:momentumbalance} to be expressed as $\bm{\nabla}_{\mathbf{x}}\cdot (\bm{\sigma} + \bm{\sigma^{\rm act}}) + \rho\mathbf{F^{\rm ext}} - \zeta \mathbf{j^\rho} = \mathbf{0}$~\cite{Omar2020}.
While derived here for ideal systems, Eq.~\eqref{eq:momentumbalance} is also general to interacting ABPs~\cite{Paliwal2018, Epstein2019} which generate an additional stress $\bm{\sigma^{\rm int}} = -\rho \langle \mathbf{xF^{int}} \rangle$.

We can now approximate the steady-state distribution of ideal ABPs as a function of ${\rm St} \equiv (m/\zeta)/\tau_R$.
In the presence of an infinite planar hard (no flux) wall at $z=0$ with a bulk density of $\rho^{\infty}$ as $z\rightarrow \infty$, ideal passive systems would recover a uniform density field $\rho(z) = \rho^{\infty}$, consistent with Boltzmann statistics.
In contrast, overdamped ABPs strongly accumulate on no-flux boundaries despite the absence of an energetic driving force for wetting the surface~\cite{Yan2015a}.
Continuing to take statistical moments of $\mathcal{P}(\mathbf{x}, \mathbf{u}, \mathbf{q})$ and closing at the nematic level (see SM~\cite{Note1} for a complete derivation), we find for $d=2$, a density profile of:
\begin{equation}
\label{eq:densityprofile}
\rho(z) = \rho^{\infty}\left(1 + \frac{D^{\rm act}}{D^B(1+{\rm St})}\exp[-z\lambda ] \right),
\end{equation}
and a polarization density of:
\begin{equation}
\label{eq:polarorderprofile}
    {\rm m}_z(z) = -\frac{\alpha \lambda D^B}{U_0}\frac{\rho^{\infty}D^{\rm act}}{\beta(D^B+D^{\rm act}) - D^{\rm act}}\exp[-z\lambda ],
\end{equation}
where  $\lambda$ is the inertia-dependent inverse screening length:
\begin{equation*}
    \lambda^2 = \frac{(d-1)\beta(1 + D^{\rm act}/D^B)}{(\alpha\gamma\delta^2)},
\end{equation*} 
where $\delta = \sqrt{\tau_RD^B}$ is a microscopic length and we have defined the following dimensionless variables:
\begin{equation*}
\alpha = 1 + \frac{D^{\rm act}}{D^B} \left(\frac{{\rm St}}{1+{\rm St}} \right),
\end{equation*}
\begin{equation*}
\beta = 1 + (d-1){\rm St},
\end{equation*}
\begin{equation*}
\gamma = \frac{1}{1+{\rm St}}\left(1 + \frac{D^{\rm act}}{D^B}\frac{(d-1){\rm St}}{2}\right).
\end{equation*}

For ${\rm St}=0$, our results exactly capture the overdamped limit~\cite{Yan2015a} while for increasing ${\rm St}$ we observe a reduction in the degree of accumulation [see Fig.~\ref{figure2}(a)] that is consistent with simulation.
These trends continue until achieving a uniform distribution of particles as ${\rm St} \rightarrow \infty$, consistent with equilibrium expectations.

The origins of this apparent effective equilibrium can be understood in the context of linear momentum conservation, which governs the dynamics of the density field.
With increasing ${\rm St}$, the decorrelation of active force and particle velocity diminishes the active stress $\bm{\sigma^{\rm act}} \sim \langle \mathbf{uF^{\rm act}} \rangle$ while the kinetic stress grows and increasingly resembles that of a passive system (i.e.,~$\bm{\sigma^K} \rightarrow -\rho k_BT^{\rm eff} \mathbf{I}$ as ${\rm St} \rightarrow \infty$, see SM~\cite{Note1}).
As a result, the steady-state momentum balance eq.~\eqref{eq:momentumbalance} becomes indistinguishable from that of a passive system, resulting in a Boltzmann distribution with an effective temperature given by the kinetic temperature $k_BT^{\rm eff} \equiv m\langle \mathbf{u} \cdot \mathbf{u} \rangle/d =  \zeta\left(D^B + D^{\rm act}\right)$~\cite{Note1}.

\begin{figure*}
	\centering
	\includegraphics[width=.95\textwidth]{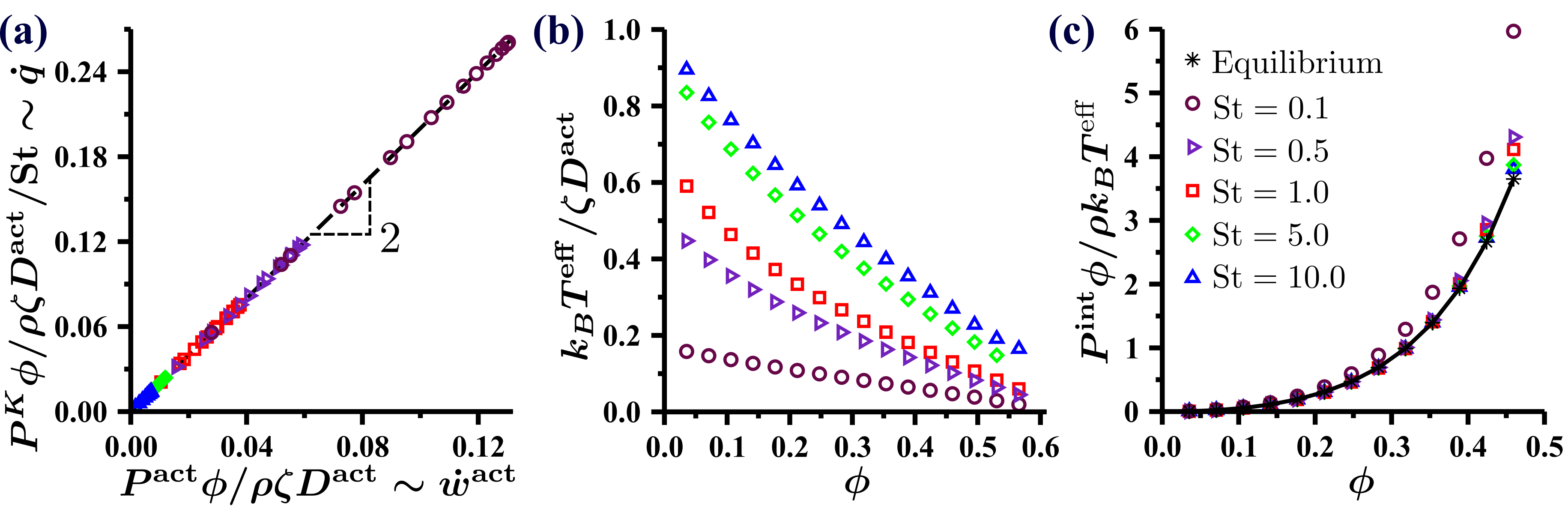}
	\caption{Active Brownian spheres with $\ell_0/\sigma = 5.0$ and in the homogeneous fluid state. (a) Parametric plot of heat and work production for all $\phi$ and ${\rm St}$ examined [see legend in panel (c)]. Dependence of (b) $k_BT^{\rm eff}$ and (c) $P^{\rm int}$ (normalized by $k_BT^{\rm eff}$) on $\phi$ and ${\rm St}$.}
	\label{figure3}
\end{figure*}

This exchange of active stress for kinetic stress (and the indistinguishably of the kinetic stress to that of a passive system) with increasing inertia, in essence, renders activity as nothing more than an energy reservoir, much like a thermal bath. 
This stress exchange is general to interacting active systems, and is in fact a consequence of the first law, which takes the form (on a per-particle basis and using the Stratanovich convention) $d\mathcal{H} = m\mathbf{u}\cdot d\mathbf{u} - \mathbf{F^C}\cdot d\mathbf{x}$. 
At steady state, the absence of average energy production results in $\frac{d}{dt}\langle \mathcal{H} \rangle = \langle \mathbf{u}\cdot (m\dot{\mathbf{u}} - \mathbf{F^C})\rangle = 0$. 
Using this and our equation-of-motion $m\mathbf{\dot{u}} = \mathbf{F^C} + \mathbf{F^{\rm source}} + \mathbf{F^{\rm drag}}$ leads to: 
\begin{equation}
    \label{eq:firstlaw}
    \underbrace{\langle -\zeta \mathbf{u\cdot u} \rangle}_{\substack{\text{dissipation} \\ {\text{rate, } \dot{q}}}} + \underbrace{\langle \mathbf{u}\cdot \mathbf{F^{\rm act}}\rangle}_{\substack{\text{active work} \\ {\text{rate, }\dot{w}^{\rm act}}}} = 0,    
\end{equation}
irrespective of if $\mathbf{F^B}$ is included in $\mathbf{F^{\rm source}}$. Equation~\eqref{eq:firstlaw} simply states that the rate of active work production is balanced by the rate of dissipation~\cite{Fodor2016, Mandal2017, Shankar2018, Nemoto2019, Tociu2019, Fodor2020, GrandPre2021, Keta2021, Goswami2022} and provides a relation between velocity-velocity and orientation-velocity correlations with deep mechanical implications.
For homogeneous and isotropic systems, this can be appreciated by taking the ratio of the 
the kinetic pressure $P^K = -\text{tr}(\bm{\sigma^K})/d = \rho m\langle \mathbf{u} \cdot \mathbf{u} \rangle/d$ to the active pressure $P^{\rm act} = -\text{tr}(\bm{\sigma^{\rm act}})/d = \frac{\tau_R}{d(d-1)}\rho\langle \mathbf{u} \cdot \mathbf{F^{\rm act}} \rangle$ which, using Eq.~\eqref{eq:firstlaw}, must result in $P^K/P^{\rm act} = (d-1){\rm St}$. 
Figure~\ref{figure3}(a) confirms this result for interacting ABPs -- active stress must be reduced in exchange for kinetic stress with ${\rm St}$.

Two crucial questions remain for interacting active systems: As $\rm{St} \rightarrow \infty$, (i) does the kinetic temperature also play the role of an effective equilibrium temperature?; and (ii) does the interaction stress $\bm{\sigma^{\rm int}}$ resemble that of passive systems at this temperature?
Figure~\ref{figure3}(b) demonstrates that the kinetic temperature of homogeneous fluids of active Brownian spheres exhibits strong dependencies on density, reaching the ideal limit $k_BT^{\rm eff} \rightarrow \zeta D^{\rm act}$ only as $\phi \rightarrow 0$ and ${\rm St} \rightarrow \infty$.
Upon normalizing the interaction pressure $P^{\rm int} = -\text{tr}(\bm{\sigma^{\rm int}})/d$ by the density-dependent kinetic temperature, we indeed observe a return to the equilibrium hard-sphere equation-of-state with increasing ${\rm St}$ for all $\phi$, as shown in Fig.~\ref{figure3}(c). 
Such a collapse would not have been possible if a density-independent energy scale had been selected and indirectly reflects the elimination of the uniquely-active contributions to $\bm{\sigma^{\rm int}}$ with ${\rm St}$. 

Physically, the origin of this equilibrium-like distribution and distinctly nonequilibrium density-dependent effective temperature can be understood by considering the various timescales present in the system.
In addition to the momentum relaxation time $\tau_M$ and active reorientation time $\tau_R$, there is a timescale associated with interactions, $\tau_{\rm int}$. 
This timescale could, for example, be the characteristic time between hard-sphere collisions. 
In the limit ${\rm St} \rightarrow \infty$, $\tau_M \gg \tau_R$ and the particle orientation -- the distinguishing feature of active matter -- ceases to play any dynamical role in the system. 
However, if $\tau_R > \tau_{\rm int}$, FDT (which requires $\tau_R$ to be much smaller than \textit{all} other timescales) remains unsatisfied.
The active force acts as a source of kinetic energy, but requires unimpeded particle motion for a duration of $\tau_R$ to generate a kinetic energy of $\zeta D^{\rm act}$. 
Collisions occurring on timescales faster than the active timescale (i.e.,~$\tau_R \gg \tau_{\rm int}$) thus reduce the kinetic energy generated by the active force. 
The increased collision rate with $\phi$ is precisely why the effective temperature decreases with concentration.

The dependence of $k_BT^{\rm eff}$ on $\phi$ suggests a coupling between temperature and local density~\cite{Fily2012, Nguyen2014, Han2017} that distinguishes ${\rm St} \rightarrow \infty$ from a true equilibrium limit.
Such a dependence may, in principle, even generate unique density fluctuations and phase transitions.
It is likely this coupling that alters the crystallization phase boundaries [cf.~Fig.~\ref{figure1}(b)] from reaching the known equilibrium values as the two phases necessarily coexist \textit{at different effective temperatures}.
This result suggests that while for \textit{spatially homogeneous} systems, inertial active matter may strongly resemble systems in thermodynamic equilibrium, situations in which strong density gradients arise (as in the case of crystallization or the presence of a strong external potential) will result in departures from equilibrium expectations.
Moreover, it is likely that even in scenarios of spatial homogeneity, the density dependence of the effective temperatures alters the distribution of local density fluctuations from equilibrium expectations.
While this is the subject of future investigation, it is likely that only the tails of the distribution are appreciably altered as significant changes to the distribution would likely be reflected in the equation-of-state [see Fig.~\ref{figure3}(c)].  

\textit{Conclusions.--} 
Despite strictly violating the FDT, translational active forces can result in particle distributions that appear to admit effective equilibrium states in the limit of large translational inertia. 
We note that \textit{rotational} inertia, in the case of ABPs, has been recently reported to accentuate MIPS by increasing the effective persistence length of particle trajectories~\cite{Sandoval2020, Lisin2022}. 
We thus emphasize that it is only in the limit of large \textit{translational} inertia that active phase transitions caused by \textit{translational} active forces are mitigated.
It would be interesting to examine the role of rotational inertia in the structures generated by chiral active matter~\cite{Zhang2022}. 
This limit can be rigorously understood in the absence of interactions, where the conversion of nonequilibrium active stress to kinetic stress renders the density distribution to be Boltzmann with the kinetic temperature serving as an effective temperature.
For the interacting systems we have studied, this limit eliminates nonequilibrium phase transitions while restoring equilibrium-like transitions.
However, a key distinction of this limit from true equilibrium is the concentration dependence of the effective temperature which leads to deviations from the anticipated equilibrium behavior.
While we have focused most of our attention on athermal systems, it is interesting to consider scenarios in which active systems are in contact with an equilibrium bath with a temperature of $k_BT$. 
In the high inertia limit, the temperature of the particles will be $k_B(T^{\rm eff}(\phi) + T)$. 
For the systems examined here,  $k_BT^{\rm eff}(\phi)$ is a monotonically decreasing function of density, and it can thus be expected that in the high density limit, the system will be \textit{truly in equilibrium} with a temperature of $k_BT$ as $T \gg T^{\rm eff}(\phi)$.
Finally, it remains to be seen if, by constructing an equation-of-state for $k_BT^{\rm eff}(\phi)$, the familiar tools of thermodynamics may be used to understand phase transitions of active systems in the large inertia limit.
Such a framework would provide a powerful tool for understanding and tuning the phase behavior of driven systems. 
In the absence of a thermodynamic framework, a recently proposed mechanical theory of nonequilibrium coexistence may be used to describe inertial phase behavior~\cite{Omar2022}.

\textit{Supplemental Material.--} 
See Supplemental Material at [URl], which includes Refs.~\cite {Risken1989, Yan2015a, Weeks1971, Omar2021, Anderson2020, Steinhardt1983, Takatori2017, delJunco2018, Plimpton1995}, for derivation and discussion of the distribution of ideal inertial ABPs, simulation and phase diagram construction details, and additional equation-of-state data for interacting ABPs in 3D.

\begin{acknowledgments}
We acknowledge helpful discussions with Cory Hargus, Kranthi Mandadapu, and Keith Burnett.
P.L.G. was supported by the U.S. Department of Energy, Office of Basic Energy Sciences, through the Chemical Sciences Division (CSD) of Lawrence Berkeley National Laboratory (LBNL), under Contract No. DE-AC02-05CH11231.
J.F.B acknowledges support by the National Science Foundation under Grant No. CBET-1803662.
We gratefully acknowledge the support of the NVIDIA Corporation for the donation of the Titan V GPU used to carry out part of this work.
The data that support the findings of this study are available from the corresponding author upon reasonable request.
\end{acknowledgments}

\end{document}

\title{Supplemental Material -- Tuning Nonequilibrium Phase Transitions with Inertia}
\author{Ahmad K. Omar}
\email{aomar@berkeley.edu}
\affiliation{Department of Materials Science and Engineering, University of California, Berkeley, California 94720, USA}
\affiliation{Materials Sciences Division, Lawrence Berkeley National Laboratory, Berkeley, California 94720, USA}
\author{Katherine Klymko}
\email{kklymko@lbl.gov}
\affiliation{NERSC, Lawrence Berkeley National Laboratory, Berkeley, California 94720, USA}
\affiliation{Computational Research Division, Lawrence Berkeley National Laboratory, Berkeley, California 94720, USA}
\author{Trevor GrandPre}
\affiliation{Department of Physics, University of California, Berkeley, California 94720, USA}
\author{Phillip L. Geissler}
\affiliation{Department of Chemistry, University of California, Berkeley, California 94720, USA}
\affiliation{Chemical Sciences Division, Lawrence Berkeley National Laboratory, Berkeley, California 94720, USA}
\author{John F. Brady}
\affiliation{Division of Chemistry and Chemical Engineering, California Institute of Technology, Pasadena, California 91125, USA}

\maketitle

\section{Fokker-Planck Analysis of Ideal Inertial ABPs}

We provide a derivation of the theory (shown in Fig.~2 in the main text) for the distribution of an ideal ABP with translational inertia.
The particle experiences active $\mathbf{F^{\rm act}} = \zeta U_0 \mathbf{q}$, drag $\mathbf{F^{\rm drag}} = -\zeta \mathbf{u}$, external (conservative)  $\mathbf{F^{\rm ext}}$ and Brownian forces $\mathbf{F^B}$ resulting in the the following translational equation-of-motion:
\begin{equation}
    \label{eq:idealeom}
    m\dot{\mathbf{u}} = \zeta U_0 \mathbf{q} - \zeta \mathbf{u} + \mathbf{F^B} + \mathbf{F^{\rm ext}}.
\end{equation}
The inclusion of Brownian forces ensures that, in the overdamped limit, the distribution of particle density is smooth and free of the singularities that occur in the non-Brownian limit (in the absence of particle interactions). 
The rotary dynamics of the particle orientation $\mathbf{q}$ follow:
\begin{equation}
    \label{eq:rotation}
    \dot{\mathbf{q}} = \bm{\Omega} \times \mathbf{q},
\end{equation}
where the angular velocity $\bm{\Omega}$ has a mean of $\bm{\Omega} = \mathbf{0}$ and a variance of $\langle \bm{\Omega}(t)\bm{\Omega}(t')\rangle = 2D_R\delta(t-t')\mathbf{I}$.

The Fokker-Planck equation describing the probability density $\mathcal{P}(\mathbf{x}, \mathbf{u}, \mathbf{q}, t)$ of finding a particle with position $\mathbf{x}$, velocity $\mathbf{u}$, and orientation $\mathbf{q}$ at time $t$ is~\cite{Risken1989}: 
\begin{equation}
\label{eq:idealfp}
\frac{\partial\mathcal{P}}{\partial t} = -\bm{\nabla}_{\mathbf{x}}\cdot\mathbf{j^x} -\bm{\nabla}_{\mathbf{u}}\cdot\mathbf{j^u}  -\bm{\nabla}_{\mathbf{q}}\cdot\mathbf{j^q},
\end{equation}
where $\mathbf{j}^{\bm{\psi}}$ and $\bm{\nabla_\psi}$ are the flux of probability and gradient operator in $\bm{\psi}$-space, respectively. 
The flux in position-space is simply due to convection $\mathbf{j^x} = \mathbf{u}\mathcal{P}$ while the rotational flux is purely diffusive $\mathbf{j^q} = -D_R\bm{\nabla}_{\mathbf{q}}\mathcal{P}$.
The flux in velocity-space contains a diffusive element stemming from the stochastic Brownian force in addition to convective contributions with:
\begin{equation}
\mathbf{j^u} = \left ( \zeta U_0 \mathbf{q}  - \zeta  \mathbf{u} + \mathbf{F^{\rm ext}} \right)\mathcal{P}/m - D^B\zeta/\tau_M\bm{\nabla}_\mathbf{u}\mathcal{P}.
\end{equation}

One can exactly determine $\mathcal{P}(\mathbf{x}, \mathbf{u}, \mathbf{q}, t)$ by numerically solving Eq.~\eqref{eq:idealfp} or, equivalently, by conducting particle-based simulations using Eq.~\eqref{eq:idealeom}. 
As our aim here is to understand the dependence of the spatial distribution of particles on inertia, we proceed analytically by defining the single-particle density $\rho(\mathbf{x}, t) = \int \mathcal{P}\,d\mathbf{u}\,d\mathbf{q}$ and determining its evolution equation [using Eq.~\eqref{eq:idealfp}] to be:
\begin{equation}
\label{eq:continuity}
\frac{\partial \rho}{\partial t} + \bm{\nabla}_\mathbf{x} \cdot \mathbf{j^\rho} = 0, 
\end{equation}
with the density flux $\mathbf{j^\rho}(\mathbf{x}, t) = \int \mathcal{P}\,\mathbf{u}\,d\mathbf{u}\,d\mathbf{q} = \rho \langle \mathbf{u} \rangle$ satisfying:
\begin{equation}
\label{eq:densityflux}
m\frac{\partial \mathbf{j^\rho}}{\partial t} + \zeta \mathbf{j^\rho} = \rho\mathbf{F^{\rm ext}} + \zeta U_0 \mathbf{m} + \bm{\nabla}_\mathbf{x} \cdot \bm{\sigma^K}.
\end{equation}
Equation~\eqref{eq:densityflux} is the statement of linear momentum conservation that was provided in the main text (with the addition of the unsteady momentum density $m\mathbf{j^\rho}$ term).
We emphasize that each term in Eq.~\eqref{eq:densityflux} is an expectation determined using the full distribution $\mathcal{P}(\mathbf{x}, \mathbf{u}, \mathbf{q}, t)$.
Determining the density flux requires solving for both the polarization density $\mathbf{m} = \rho \langle \mathbf{q} \rangle$ and the kinetic stress $\bm{\sigma^K} = -\rho m \langle \mathbf{uu} \rangle$ (where $\rho \langle \bm{\psi} \rangle =  \int \mathcal{P}\,\bm{\psi}\,d\mathbf{u}\,d\mathbf{q}$).
The evolution equation for the stress is given by:
\begin{equation}
\label{eq:kstress}
\frac{\partial \bm{\sigma^K}}{\partial t} + \frac{2}{\tau_M}\bm{\sigma^K} = -\frac{2D^B\zeta}{\tau_M} \rho \mathbf{I} - \left(\mathbf{F^{\rm ext}}\mathbf{j^\rho} + \mathbf{j^\rho}\mathbf{F^{\rm ext}} \right ) - U_0 \zeta \left(\mathbf{j^m} + (\mathbf{j^m)^\top} \right),
\end{equation}
where $\mathbf{j^m} = \int \mathcal{P}\,\mathbf{uq}\,d\mathbf{u}\, d\mathbf{q}$ is the polarization flux.
Before proceeding to the polarization conservation equations, it is worth examining the structure of Eq.~\eqref{eq:kstress}.
At steady state, a passive system would recover a kinetic stress of $\bm{\sigma^K}=-\rho \zeta D^B \mathbf{I} = -\rho k_BT\mathbf{I}$ that is characterized by a constant temperature $k_BT$.
The active contribution to the kinetic stress [the last term in Eq.~\eqref{eq:kstress}] scales as $\sim U_0m\mathbf{j^m} = U_0m\rho\langle \mathbf{uq} \rangle$ and, as a result, the kinetic energy arising from activity may vary spatially in response to changes in polarization flux. 
Even for ideal ABPs, the kinetic temperature is not guaranteed to be spatially constant.

The equations analogous to Eqs.~\eqref{eq:continuity}, \eqref{eq:densityflux}, and \eqref{eq:kstress} for the polarization field are, respectively:
\begin{equation}
\label{eq:mcontinuity}
\frac{\partial \mathbf{m}}{\partial t} + \bm{\nabla_{\mathbf{x}}}\cdot \mathbf{j^m} + (d-1)D_R\mathbf{m} = 0, 
\end{equation}
\begin{equation}
\label{eq:mflux}
m\frac{\partial \mathbf{j^m}}{\partial t} + m\left(\frac{1}{\tau_M} + \frac{d-1}{\tau_R}\right) \mathbf{j^m} = \mathbf{F^{\rm ext}}\mathbf{m} + \zeta U_0 \mathbf{\tilde{Q}} + \bm{\nabla}_\mathbf{x} \cdot \mathbf{D^m}, 
\end{equation}
\begin{equation}
\label{eq:mkstress}
\frac{\partial {\rm D}_{ijk}^{\rm m}}{\partial t} + \left(\frac{2}{\tau_M} + \frac{2}{\tau_R}\right) {\rm D}_{ijk}^{\rm m} = -\frac{2D^B\zeta}{\tau_M} \delta_{ij} {\rm m}_k - \left({\rm F}_i^{\rm ext} {\rm j}_{jk}^{\rm m} + {\rm F}_j^{\rm ext} {\rm j}_{ik}^{\rm m} \right ) - U_0 \zeta \left( {\rm j}_{ijk}^{\tilde{\rm Q}} + {\rm j}_{jik}^{\tilde{\rm Q}} \right),
\end{equation}
where $\mathbf{\tilde{Q}}=\rho\langle \mathbf{qq} \rangle$ is the nematic tensor, and $\mathbf{D^m} = -\rho m\langle \mathbf{uuq} \rangle$ drives the diffusion of the polarization flux. 
Note that we have switched to Einstein notation (e.g.,~$\mathbf{j^m} \rightarrow  {\rm j}_{ik}^{\rm m}$) in Eq.~\eqref{eq:mkstress}.

Equations~\eqref{eq:continuity}-\eqref{eq:mkstress} do not constitute a closed set of equations -- conservation and constitutive equations describing the nematic field and flux ${\rm j}_{ijk}^{\tilde{\rm Q}}$ (which will further depend on higher-order orientational moments) are required.
To close the hierarchy of equations, we assume the nematic field to be isotropic $\mathbf{\tilde{Q}} = \rho \mathbf{I}/d + \mathbf{Q} \approx \rho \mathbf{I}/d$, a familiar closure that in many cases comes at little quantitative expense~\cite{Yan2015a}. 
Using this closure, the nematic flux can now be expressed as ${\rm j}_{ijk}^{\tilde{\rm Q}} = \langle {\rm u}_i {\rm q}_j {\rm q}_k \rangle = \langle {\rm u}_i \delta_{jk}/d \rangle + \langle {\rm u}_i ({\rm q}_j {\rm q}_k - \delta_{jk}/d)\rangle\approx \delta_{jk}{\rm j}_i^\rho/d$.
Additionally, in the flux expressions [Eqs.~\eqref{eq:densityflux} and \eqref{eq:mflux}], we only retain terms up to a single spatial gradient, ensuring that our field equations [Eqs.~\eqref{eq:continuity} and \eqref{eq:mcontinuity}] are both second order with respect to spatial gradients. 

We can now determine the stationary density distribution near a hard wall at $z=0$ in the absence of external forces $\mathbf{F^{\rm ext}} = \mathbf{0}$.
Conservation of density and polarization result in $\bm{\nabla} \cdot \mathbf{j^\rho} = 0$ and $\mathbf{m} = \tau_R\bm{\nabla} \cdot \mathbf{j^m}/(d-1)$, respectively. 
The density and polarization fluxes satisfy:
\begin{equation}
\mathbf{j^\rho} = U_0 \mathbf{m} + \frac{1}{\zeta}\bm{\nabla}_\mathbf{x} \cdot \bm{\sigma^K},
\end{equation}
\begin{equation}
\mathbf{j^m} = \frac{1}{1+(d-1){\rm St}}\left( \frac{U_0\rho}{d}\mathbf{I} + \frac{1}{\zeta}\bm{\nabla}_{\mathbf{x}}\cdot \mathbf{D^m}\right),
\end{equation}
where we have invoked our $\mathbf{Q} = \mathbf{0}$ closure in $\mathbf{j^m}$.
Using our expression for the nematic flux, we obtain:
\begin{equation}
{\rm D}_{ijk}^{\rm m} = -\zeta D^B \gamma \delta_{ij} {\rm m}_k,
\end{equation}
where we have defined:
\begin{equation*}
\gamma = \frac{1}{1+{\rm St}}\left(1 + \frac{D^{\rm act}}{D^B}\frac{(d-1){\rm St}}{2}\right),
\end{equation*}
with $D^{\rm act} = \frac{U_0^2\tau_R}{d(d-1)}$.
We have neglected the gradient terms that arise from the nematic flux (i.e.,~${\rm j}_{ijk}^{\tilde{\rm Q}} \approx \delta_{jk}{\rm j}_i^\rho/d \approx \delta_{jk}U_0{\rm m_i}/d$), consistent with our aim of retaining up to second order spatial gradients at the field level (or first order at the flux level). The polarization flux can now be expressed as: 
\begin{equation}
\mathbf{j^m} = \frac{1}{1+(d-1){\rm St}}\left( \frac{U_0\rho}{d}\mathbf{I} - \gamma D^B\bm{\nabla}_{\mathbf{x}}\mathbf{m}\right),
\end{equation}
Finally, the kinetic stress is found to be:
\begin{equation}
\bm{\sigma^K} = -\zeta \left(D^B + \frac{D^{\rm act}(d-1) {\rm St}}{1 + (d-1){\rm St}}\right)\rho \mathbf{I}.
\end{equation}
We emphasize that in deriving this stress, we omitted the gradient contributions to $\mathbf{j^m}$ in addition to any nematic alignment $\mathbf{Q}$ and, as a result, activity \textit{uniformly} increases the kinetic temperature of the system.
The inclusion of these omitted effects could result in a nonuniform kinetic temperature in the presence of density gradients and/or the nematic alignment of particles. 
The deviation from a uniform temperature due to these higher-order effects are separate from what was observed for interacting ABPs, which have a density-dependent kinetic temperature stemming from particle interactions (see Fig.~3(b) in the main text). 

The derived conservation and constitutive equations along with the boundary conditions of no-flux at the wall surface (${\rm j}_z^\rho = {\rm j}_{zz}^m = 0$ at $z=0$) and uniform orientations $\mathbf{m}^{\infty} = \mathbf{0}$ and bulk density $\rho^{\infty}$ as $z\rightarrow \infty$ results in the following equations for the density and polarization fields:
\begin{equation}
\label{eq:densityfield}
\rho(z) = \rho^{\infty}\left(1 + \frac{D^{\rm act}}{D^B\beta}\exp[-z\lambda ] \right),
\end{equation}
\begin{equation}
\label{eq:mfield}
    {\rm m}_z(z) = -\frac{\alpha \lambda D^B}{U_0}\frac{\rho^{\infty}D^{\rm act}}{\beta(D^B+D^{\rm act}) - D^{\rm act}}\exp[-z\lambda ],
\end{equation}
where the screening length is:
\begin{equation}
\label{eq:lambda}
    \lambda^2 = \frac{(d-1)\beta(1 + D^{\rm act}/D^B)}{(\alpha \gamma \delta^2)},
\end{equation} 
and where we have defined constants $\alpha$ and $\beta$:
\begin{equation*}
\alpha = 1 + \frac{D^{\rm act}}{D^B} \left(\frac{{\rm St}}{1+{\rm St}} \right),
\end{equation*}
\begin{equation*}
\beta = 1 + (d-1){\rm St}.
\end{equation*}

\begin{figure}
	\centering
	\includegraphics[width=.4\textwidth]{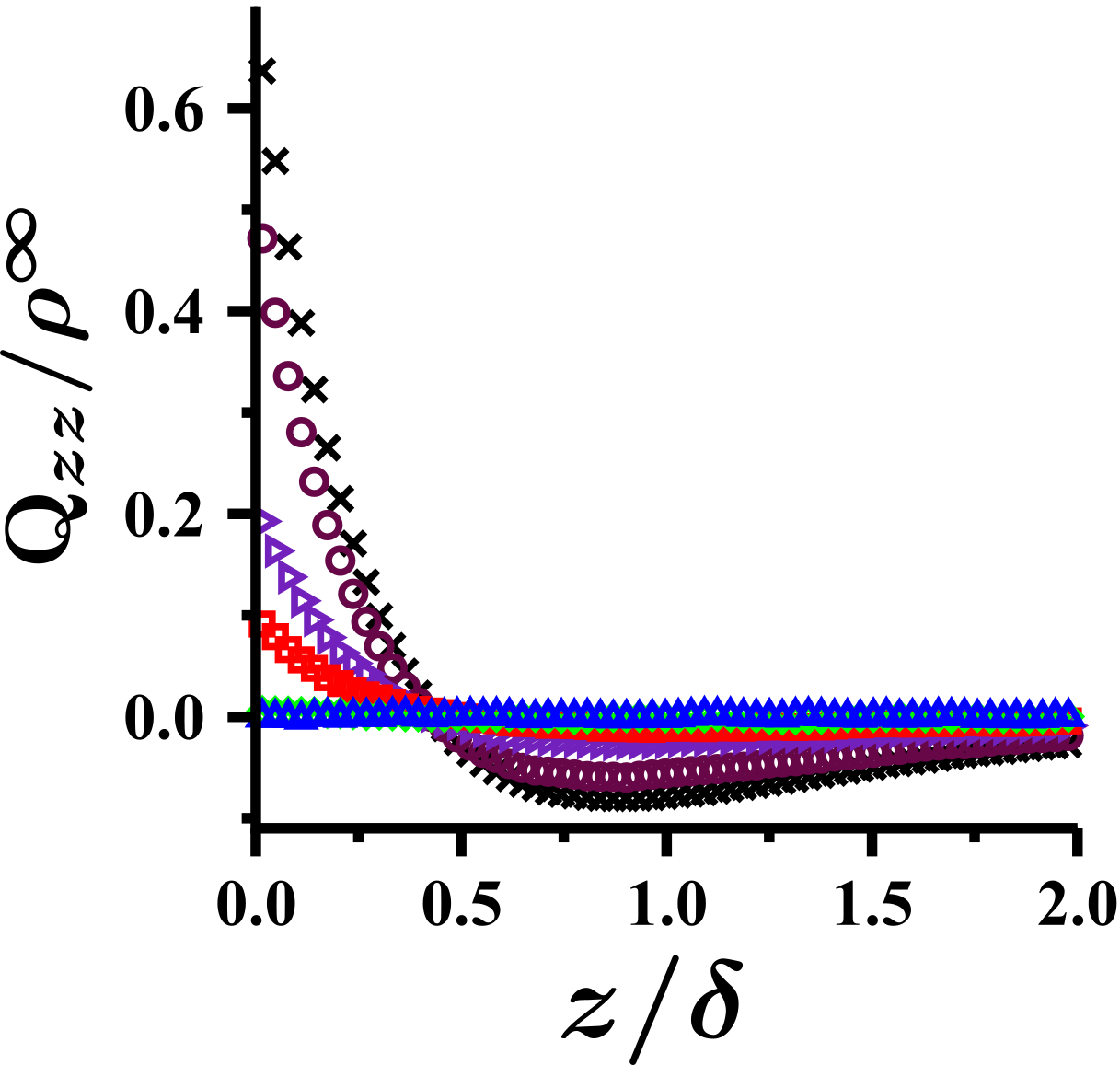}
	\caption{\protect\small{{{\rm St} dependence of the nematic field (taken to be ${\rm Q}_{zz} = 0$ in our theory) from the same simulations used to generate Fig.~2 in the main text.}}}
	\label{qfield}
\end{figure}

Equations~\eqref{eq:densityfield} and \eqref{eq:mfield} for the density and polarization fields are shown in comparison to the results from particle-based simulations in Fig.~2 in the main text.
Our results, which only contain gradients up to second order in the field conservation equations and neglect the nematic alignment of particles, capture the essential trends with changing {\rm St}.
Namely, we find a reduction in the degree of particle accumulation and polarization with increasing inertia that is consistent with simulation.
From simulations, we are able to see that higher-order orientational moments also become increasingly isotropic with increasing inertia, as shown for the nematic field in Fig.~\ref{qfield}.

Finally, in the presence of an external field $\mathbf{F^{\rm ext}}$ it is straightforward to show that in the ${\rm St} \rightarrow \infty$ limit, the solution to our coupled equations is simply:
\begin{equation}
    \label{eq:equilibrium}
    \rho(\mathbf{x}) = \exp[-V^{\rm ext}(\mathbf{x})/k_BT^{\rm eff}]
\end{equation}
where $\mathbf{F^{\rm ext}} = -\bm{\nabla}_\mathbf{x}V^{\rm ext}$ and $k_BT^{\rm eff} = \zeta (D^B + D^{\rm act})$. 
Importantly, the density gradient terms (that arise from $\mathbf{j^m}$) neglected in $\bm{\sigma^K}$ could result in deviations from precisely Boltzmann statistics and, as a result, departures from Eq.~\eqref{eq:equilibrium} might be expected in the presence of strong external fields.

\section{Simulation and Calculation Details}

\subsection{Ideal ABP Simulations}

For the 2D ideal ABP simulations shown in Fig.~2 in the main text, the no-flux boundary condition was enforced by implementing a reflective boundary condition at the wall surface. 
The reversal of the particle velocity upon colliding with the wall ensures that the collision is perfectly elastic.
In contrast to our theory, which was for a semi-infinite system, our simulations contained walls on opposite sides of the domain.
Care was taken to ensure that the wall separation was significant enough such that the boundaries did not interfere with each other.

\subsection{Interacting Active Systems}

For all of the interacting systems examined in this work, we take the interparticle force to result from a Weeks-Chandler-Anderson potential~\cite{Weeks1971} (characterized by a Lennard-Jones diameter $\sigma$ and energy $\varepsilon$) $\mathbf{F^{\rm int}}(\mathbf{x}_{ij}) = -\boldsymbol{\nabla}\varepsilon u^\text{WCA}(r; \sigma)$ where $\mathbf{x}_{ij}$ is the particle separation (with magnitude $r = |\mathbf{x}_{ij}|$) and the reduced potential $u^\text{WCA}$ has the following form~\cite{Weeks1971}:
\begin{equation}
u^\text{WCA}(r; \sigma) =
 \begin{cases}
   4\left[\left(\frac{\sigma}{r}\right)^{12} - \left(\frac{\sigma}{r}\right)^{6}\right] + 1, & r \le 2^{1/6}\sigma \\
    0, & r > 2^{1/6}\sigma.  \\
    \end{cases}
\end{equation}
Upon selecting $\zeta U_0$ and $\sigma$ as the respective units of force and length, the resulting dimensionless force $\mathbf{\overline{F}^{\rm int}}_{ij}(\overline{\mathbf{x}}_{ij}; \mathcal{S}) = -\mathcal{S}\overline{\boldsymbol{\nabla}} u^\text{WCA}(\overline{r})$ (where $\overline{\boldsymbol{\nabla}} = \sigma \boldsymbol{\nabla}$ is the dimensionless gradient operator) is entirely characterized by the stiffness $\mathcal{S}$.

\begin{figure*}
	\centering
	\includegraphics[width=.75\textwidth]{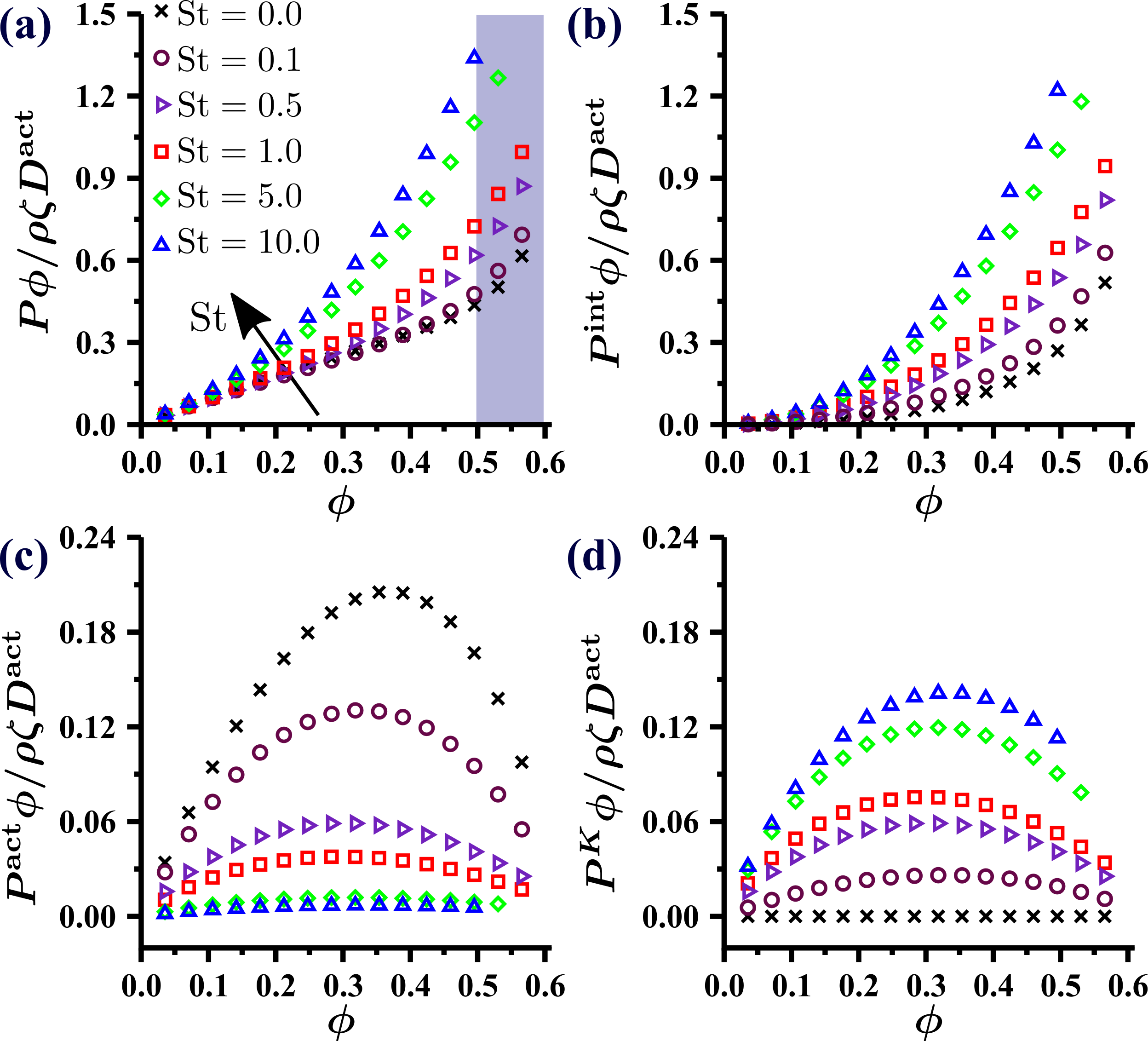}
	\caption{\protect\small{{${\rm St}$ and $\phi$ dependencies of the (a) total, (b) interparticle, (c) active (or swim), and (d) kinetic pressures for active Brownian spheres with $\ell_0/\sigma = 5.0$.}}}
	\label{pressure}
\end{figure*}

\subsection{3D Interacting, Athermal ABPs}

In the overdamped limit, the three dimensionless parameters that fully describe the system state are $\ell_0/\sigma$, volume fraction $\phi$ and the stiffness parameter $\mathcal{S}$. 
The introduction of inertia adds a fourth parameter, the Stokes number, with ${\rm St} = (m/\zeta)/\tau_R$ where $m$ is the particle mass. 
$\mathcal{S} = \varepsilon/(\zeta U_0 \sigma)$ plays a determining role in recovering hard-sphere packing statistics.
A choice of $\mathcal{S} = 50.0$ (as was used in Ref.~\cite{Omar2021}) when the particle dynamics are overdamped ensures that the active force cannot generate overlaps within a pair separation $d_{\rm ps}$ of $d_{\rm ps}/(2^{1/6}\sigma) = 0.9997$. 
While inertia can, in principle, result in particle overlaps within $d_{\rm ps}$ that are otherwise not possible in the overdamped limit, for the {\rm St} examined in this work, particle overlaps remained insignificant with $\mathcal{S} = 50.0$.
Achieving effective hard-sphere statistics allows us to take $2^{1/6}\sigma$ as the particle diameter and define the volume fraction as $\phi = \rho \pi (2^{1/6} \sigma)^3/6$ where $\rho = N/V$ is the number density and $V$ is the system volume.
All active Brownian sphere simulations were conducted with a minimum of $54000$ particles using \texttt{HOOMD-blue}~\cite{Anderson2020}.

The inertia dependence of the MIPS phase boundary presented in Fig.~1(a) was determined for ABPs with a runlength of $\ell_0/\sigma = 50.0$, ensuring that the simulation is well within the MIPS phase boundary in the overdamped limit ${\rm St} = 0$.
We extracted the coexisting densities by conducting simulations in skewed periodic domains which bias the orientation of a liquid-gas interface and allows us to determine the density profile, following the procedure outlined in Ref.~\cite{Omar2021}.

To obtain the crystallization coexistence curve in Fig.~1(b) (now for $\ell_0/\sigma = 5.0$), rather than introducing a crystal seed in our simulation, we varied the total simulation density (as a function of ${\rm St}$) until a spontaneous nucleation was observed.
The simulated densities that resulted in these nucleation events for ${\rm St} = 0.0, 0.1, 1.0, 2.0, 3.0, 5.0,$ and $10.0$ were $\phi = 0.64, 0.64, 0.64, 0.60, 0.60, 0.60,$ and $0.57$, respectively.
These densities were reached by initially preparing the system at a low volume fraction ($\phi=0.2$) followed by an isotropic compression to the target density, as described in Ref.~\cite{Omar2021}. 
All simulations were run for a minimum duration of $25000~\sigma/U_0$.
We then computed the distributions of the local densities, taking the local maxima to correspond to the coexisting densities.
When the bimodal character of the density distribution is unclear, we compute the local density distribution of ``ordered" and ``disordered" particles separately and independently compute their averages to obtain the coexisting densities.
To determine the local order of a particle, we compute the Steinhardt-Nelson-Ronchetti~\cite{Steinhardt1983} order parameter defined as:
\begin{equation}
\label{eq:localq12pt1}
q_l(i) = \biggl (\frac{4\pi}{2l+1} \sum_{m=-l}^{l}|\langle Y_{lm}\rangle|^2 \biggr)^{1/2},
\end{equation}
where $\langle Y_{lm} \rangle$ is the average spherical harmonics of the bond angles formed between particle $i$ and its nearest neighbors. 
We define crystalline particles as those with $q_{12} > 0.50$ while fluid-like particles are defined as those with $q_{12} < 0.35$

\begin{figure}
	\centering
	\includegraphics[width=.4\textwidth]{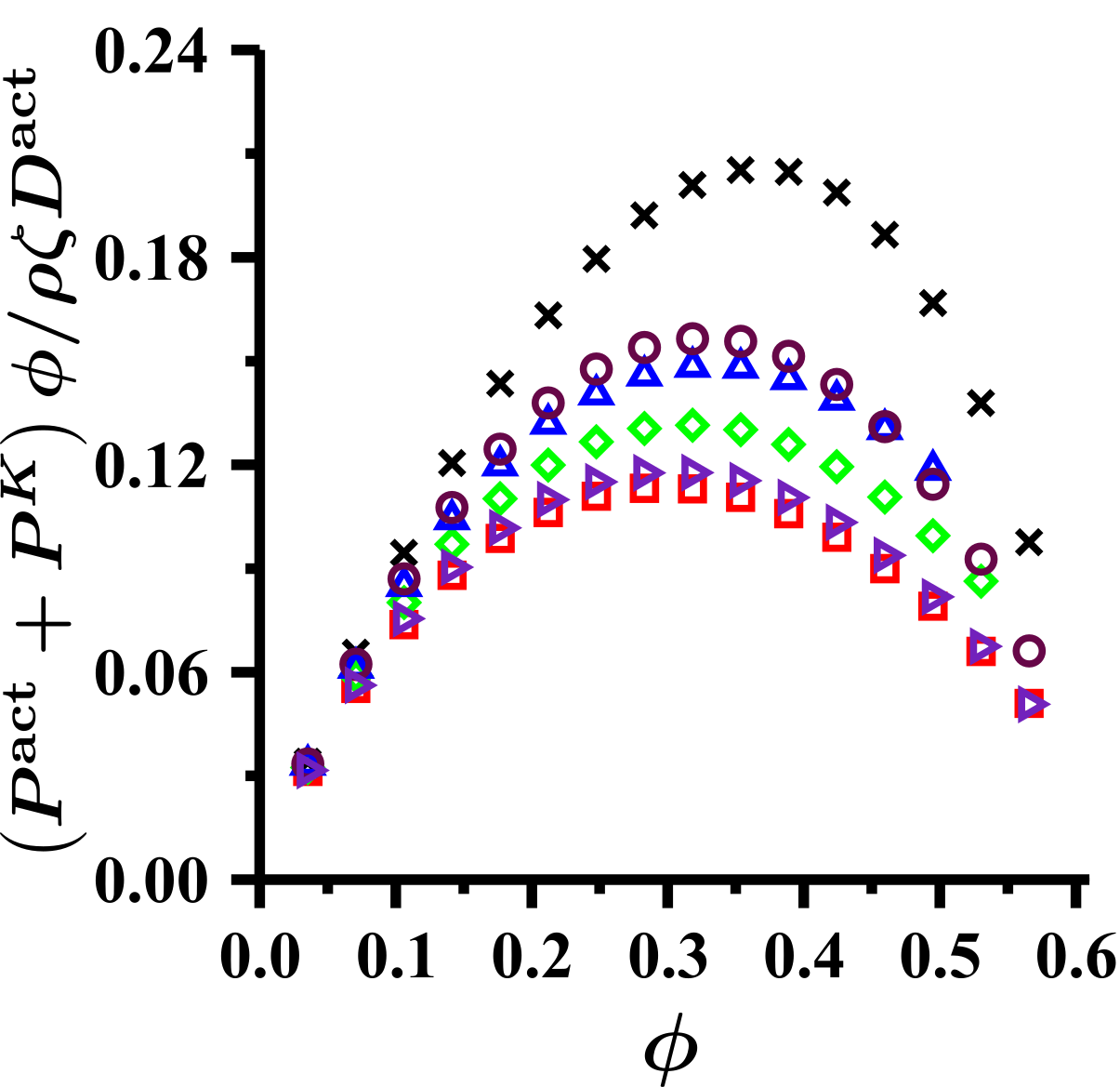}
	\caption{\protect\small{{${\rm St}$ and $\phi$ dependencies of the sum of the active and kinetic pressures for active Brownian spheres with $\ell_0/\sigma = 5.0$. }}}
	\label{swimpluskinetic}
\end{figure}

The ${\rm St}$ dependence of the total pressure $P = P^{\rm act} + P^K + P^{\rm int}$ for a homogeneous fluids of active Brownian spheres is shown in Fig.~\ref{pressure}.
For large ${\rm St}$, homogeneous fluids can spontaneously crystallize in the region of density shaded in Fig.~\ref{pressure}(a). 
These points are subsequently omitted to ensure that the equation-of-state presented is for acive fluids.
We note that in Ref.~\cite{Takatori2017}, it was shown analytically that the sum of the active and kinetic pressures is independent of ${\rm St}$ and it was further argued that this holds for interacting  particles as well.
For our active Brownian spheres, this does not hold, as there is a nontrivial ${\rm St}$ dependence to $P^K + P^{\rm act}$ as shown in Fig.~\ref{swimpluskinetic}.
Equation~(4) in the main text in fact shows that it is the ratio of these two pressures that is independent of density and linearly related to ${\rm St}$ for interacting systems.

\subsection{2D Oscillating/Passive Disk Mixture}

We conduct identical simulations to those performed in Ref.~\cite{delJunco2018}.
In contrast to our active Brownian spheres, here the particles are relatively soft with $\varepsilon/k_BT = 1$.
All simulations were conducted with $10124$ particles and for a duration of $1000~\sigma^2/D^B$ using \texttt{LAMMPS}~\cite{Plimpton1995}.
To characterize the influence of inertia on pattern formation, we compute the average length scale associated with the patterns formed by first computing the static structure factor of the $N_A$ active particles:
\begin{equation}
\label{eq:Sk}
S(\mathbf{k}) = \frac{1}{N_A}\biggl\langle\biggl|\sum_{j=1}^{N_A}\exp(\mathrm{i}\mathbf{k}\cdot \mathbf{x}_j)\biggr|^2\biggr\rangle.
\end{equation}
We then associate a length scale with the structural correlations by taking the first moment of the structure factor with respect to ${\rm k} = |\mathbf{k}|$:
\begin{equation}
\label{eq:L}
\langle L \rangle^{-1} = \frac{\int_{{\rm k}_{\rm min}}^{{\rm k}_{\rm max}} S({\rm k})\,{\rm k}\,d{\rm k}}{\int_{{\rm k}_{\rm min}}^{{\rm k}_{\rm max}} S({\rm k})\,d{\rm k}},
\end{equation}
where ${\rm k}_{\rm min}$ is set by the simulation domain size and ${\rm k}_{\rm max} = \pi/\sigma$ was selected as a cutoff.